# Revealing nanoscale structural phase separation in La$_3$Ni$_2$O$_{7-\delta}$ single crystal via scanning near-field optical microscopy

Xiaoxiang Zhou[1,†], Weihong He[1,2,†], Kaipeng Ni[3,†], Mengwu Huo[4], Deyuan Hu[4], Yinghao Zhu[5,6], Enkang Zhang[5,6], Zhicheng Jiang[2], Shuaikang Zhang[7], Shiwu Su[7], Juan Jiang[1,7], Yajun Yan[1,7], Yilin Wang[1,7], Dawei Shen[2], Xue Liu[3]*, Jun Zhao[1,5,6], Meng Wang[4], Zengyi Du[1]*, Donglai Feng[1,2,7,8]*

[1] *Hefei National Laboratory, Hefei, Anhui 230088, China*

[2] *National Synchrotron Radiation Laboratory and School of Nuclear Science and Technology, University of Science and Technology of China, Hefei, 230026, China*

[3] *Key laboratory of Magnetic Functional Materials and Devices of Anhui Province, Center of Free Electron Laser & High Magnetic Field, Anhui University, Hefei 230601, China*

[4] *Guangdong Provincial Key Laboratory of Magnetoelectric Physics and Devices, School of Physics, Sun Yat-Sen University, Guangzhou, Guangdong 510275, China*

[5] *State Key Laboratory of Surface Physics, and Department of Physics, Fudan University, Shanghai 200433, China*

[6] *Shanghai Research Center for Quantum Sciences, Shanghai 201315, China*

[7] *School of Emerging Technology, University of Science and Technology of China, Hefei 230026, China*

[8] *New Cornerstone Science Laboratory, University of Science and Technology of China, Hefei, 230026, China*

[†] Those authors contribute equally to this work

*Correspondence emails: lxue@ahu.edu.cn, duzengyi@ustc.edu.cn, dlfeng@ustc.edu.cn

The discovery of superconductivity in La$_3$Ni$_2$O$_{7-\delta}$ under high pressure, with an onset critical temperature($T_c$) around 80 K, has sparked significant interest in the superconducting phases of Ruddlesden-Popper nickelates, La$_{n+1}$Ni$_n$O$_{3n+1}$ (n=2, 3). While La$_4$Ni$_3$O$_{10}$ exhibits nearly 100% superconductivity with $T_c\sim$30K under high pressure, magnetic susceptibility studies on La$_3$Ni$_2$O$_{7-\delta}$, however, reveal a more complex picture, indicating either filamentary superconductivity or that approximately 50% of crystal phase becomes superconducting in polycrystalline samples. In this study, we employed scattering-type scanning near-field optical microscopy (SNOM) to visualize nanoscale structural phase separation in La$_3$Ni$_2$O$_{7-\delta}$, identifying enhanced optical conductivity with stripes approximately 183 nm wide. These stripes run diagonally with respect to the Ni-O-Ni bond directions in the a-b plane, ruling out the possibility that they arise from impurity phases, like the '1313', '214' or '4310' structures. The dark regions and bright stripes exhibit optical conductivities ~ 22% and 29% of gold's, respectively. Additionally, we find that the bright stripes constitutes about 38% of the total field of view, while the remainder consists of dark regions and the transitional region between dark regions and bright stripes. Our results suggest that optical conductivity stripes originate from nanoscale structural phase separation. In contrast, La$_4$Ni$_3$O$_{10}$ exhibits uniform and higher optical conductivity with no observable evidence of phase separation. Thus, our study represents a pioneering effort to directly image nanoscale phase separation in La$_{n+1}$Ni$_n$O$_{3n+1}$ (n=2,3) nickelates. This observation could provide crucial insights into the factors that limit the superconducting volume fraction of La$_3$Ni$_2$O$_{7-\delta}$, highlighting SNOM as a powerful probe for exploring nanoscale low-energy

**physics in correlated quantum materials.**

Decades of efforts in searching for superconductivity in nickelates have been fulfilled with[1-3] the successful synthesis of superconducting nickelate thin films of $Nd_{0.8}Sr_{0.2}NiO_2$[4-6]. These thin films exhibit a transition temperature ($T_c$) of around 15 K, which increases to over 31 K under external pressure[7]. Superconductivity has also been observed under high pressure in single crystal Ruddlesden-Popper (RP) phases of $La_{n+1}Ni_nO_{3n+1}$ family, particularly for n = 2 and 3[8-16]. Recently, the appearance of superconductivity with $T_c$ larger than 40 K has been reported in bulk sample $La_3Ni_2O_7$ and in thin films of parent and Pr-doped $La_3Ni_2O_7$ at ambient pressure[17-19], indicating that nickelates are likely to serve as a new platform for researching high $T_c$ superconductivity at ambient pressure. Notably, the bilayer compound $La_3Ni_2O_{7-\delta}$ demonstrates an onset $T_c$ exceeding the liquid nitrogen temperature (77 K)[8-12] While nearly 100% superconductivity has been experimentally observed in the tri-layer compound $La_4Ni_3O_{10}$ ($T_c \sim 30$ K)[13], the primary superconducting phase of $La_3Ni_2O_{7-\delta}$ remains elusive as recent studies on single crystals suggest filamentary superconductivity[20-21] while research on $La_2PrNi_2O_{7-\delta}$ polycrystalline samples indicates about 57% of crystals is superconducting[22].

X-ray and neutron diffraction refinements reveal that, at ambient pressure, both $La_3Ni_2O_{7-\delta}$ and $La_4Ni_3O_{10}$ crystallize in orthorhombic structures. Specifically, $La_3Ni_2O_{7-\delta}$ adopts either the *Amam* or *Fmmm* structure[8,12-13,21-24], whereas $La_4Ni_3O_{10}$ is in the $P2_1/a$ structure[13-16]. Both compounds experience a progressive reduction in lattice constants and unit cell volumes upon applying high pressure, accompanied by a structural transition at ~12 GPa from orthorhombic to tetragonal (*I4/mmm*) structure. In the related $La_2PrNi_2O_{7-\delta}$ polycrystalline samples, partial substitution of La by smaller Pr atoms, effectively applying chemical pressure, improves the purity of the '327' phase, resulting in a bulk superconducting phase close to 57% at 20 GPa in *I4/mmm* structure[22]. These observations support the proposition that the tetragonal *I4/mmm* structure is the primary superconducting phase in both $La_3Ni_2O_{7-\delta}$ and $La_4Ni_3O_{10}$ at low temperatures.

To elucidate the reduced superconducting fraction, it has been proposed that oxygen deficiencies of crystal structure, especially the oxygen deficiency between two adjacent Ni-O octahedra, have profound effect on both superconductivity and electronic structure[22-23,25-26]. The low energy physics of $La_3Ni_2O_{7-\delta}$ is dominated by interlayer coupling[27-35], in which electrons virtually hop between the Ni-$3d_{z^2}$ orbitals in adjacent Ni-O octahedra through the apical oxygen $2p_z$ orbitals. Resonant inelastic X-ray scattering (RIXS) study has revealed that the interlayer exchange interaction plays a crucial role in determining electronic and magnetic excitations of $La_3Ni_2O_{7-\delta}$[36]. Electron counting reveals that the $3d_{z^2}$ bands in $La_3Ni_2O_{7-\delta}$ are close to half-filling, exhibiting strong correlations as evidenced by angle-resolved photoemission spectroscopy (ARPES)[37] and infrared optical spectroscopy[38]. The $3d_{x^2-y^2}$ orbitals are near quarter-filling, which leads to metallic behavior in the stoichiometric $La_3Ni_2O_7$ ('327') phase, consistent with transport measurements in samples with minimal oxygen deficiencies ($\delta \sim 0$)[39-40]. However, recent multislice electron ptychography (MEP) studies uncovered significant inhomogeneity in the oxygen content of $La_3Ni_2O_{7-\delta}$, with oxygen vacancies leading to a nonstoichiometric $\delta$ as high as 0.34[41]. In $La_3Ni_2O_{7-\delta}$, as the oxygen deficiency ($\delta$) increases, electron scattering gets enhanced, leading to the suppressed local conductivity and a transition toward insulating behavior, as observed in transport measurements[39-40]. In extreme cases, the absence of inner oxygens can locally transform the '327' phase into an insulating '326' phase or the '4310' into the '438' phase[42-46]. Therefore, it is crucial to determine whether this ubiquitous oxygen inhomogeneity is responsible for the reduced superconducting volume fraction as observed in $La_3Ni_2O_{7-\delta}$.

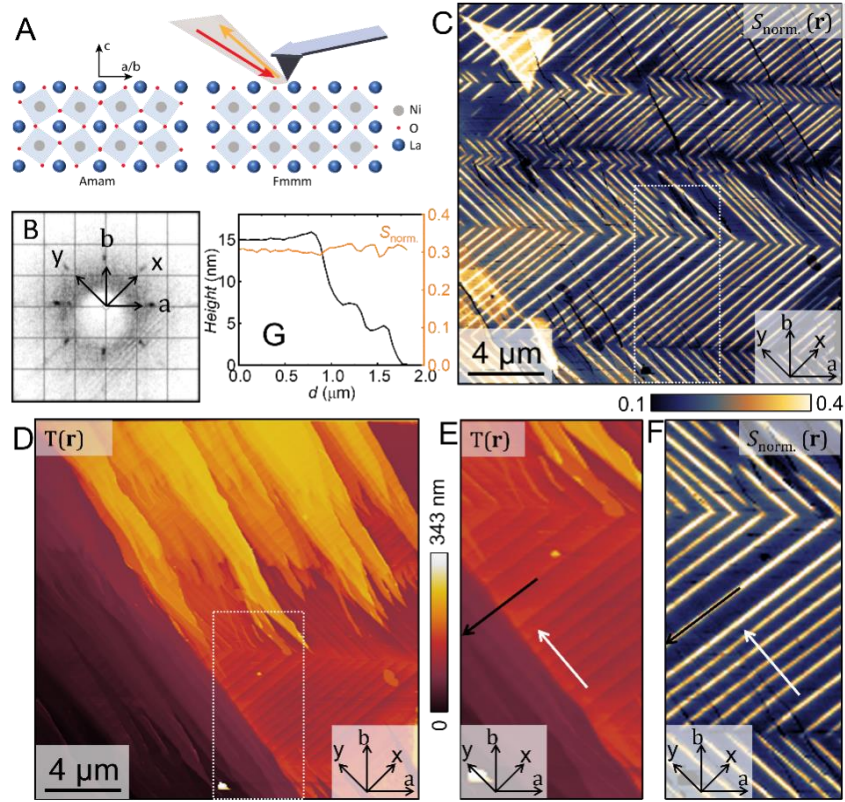

Figure. 1| **A.** Schematic of the experimental SNOM setup and crystal models for $La_3Ni_2O_{7-\delta}$ in *Amam* and *Fmmm* structures respectively. Note that the Ni-O-Ni angle between two adjacent octahedra is ~168° in the *Amam* structure and 180° in the *Fmmm* structure. The SNOM study is conducted within the a-b plane of $La_{n+1}Ni_nO_{3n+1}$ (n = 2,3) samples. **B**. The Laue pattern measured on the $La_3Ni_2O_{7-\delta}$ sample. Here, arrows a and b represent Ni-O-Ni crystal axes, while x and y denote the directions of optical conductivity stripes. **C**. The near-field signal map $S(\mathbf{r})$ acquired on $La_3Ni_2O_{7-\delta}$ at ambient pressure. Clear nanoscale conductivity stripes are observed, and these conductivity stripes form domains, with the orientation of the stripes rotating by 90 degrees between neighboring domains. Near-field signal $S(\mathbf{r})$ of $La_3Ni_2O_{7-\delta}$ was normalized against the signal measured on gold. **D.** The AFM topography $T(\mathbf{r})$ measured simultaneously with $S(\mathbf{r})$ in Fig. C within the same field of view (FOV). Step edges can be identified running from top-left to bottom-right corners. Structural change can also be identified from the topography following the same tracks of optical conductivity stirpes in Fig. 1C. Those correspondence structural change are running top-right to bottom-left corners. **E, F**. Zoomed in topography $T(\mathbf{r})$ and near-field map $S(\mathbf{r})$ cropped from white dashed rectangle in Figs. CD. **G**. Line profiles along black arrows in Figs. EF, crossing a step edge with a height of ~15 nm without disruption the optical conductivity stripes.

In this study, we investigated the nanoscale low-energy physics of $La_3Ni_2O_{7-\delta}$ and $La_4Ni_3O_{10}$ by imaging the local optical conductivity using the scattering-type scanning near-field optical microscopy (SNOM)[47-50]. The light scattered from an atomic force microscopy

(AFM) tip (as illustrated in Fig. 1A), encodes detailed information about the local optical properties of the sample. The spatial resolution of the near-field signal is set by the AFM tip radius ($r \sim 10\text{–}20$ nm), which is well below the Abbe diffraction limit for the 8.0 μm-wavelength incident light. Similar to optical conductivity measured in 'far field' infrared optical experiment[50-52], the collected near-field signal, $S$, also yields bulk optical conductivity, which can be approximately described by $S \propto \beta = \frac{(\varepsilon_s - 1)}{(\varepsilon_s + 1)}$, where $\varepsilon_s \approx \varepsilon' - i\frac{\sigma}{\omega}$ represents the sample's dielectric constant, and $\sigma$ denotes optical conductivity. Typically, higher near-field signal $S$ values resulting from the Drude response are expected for metals, while lower values correspond to less conductive materials and zero indicative of insulators[53-56].

We obtained the $La_3Ni_2O_{7-\delta}$ and $La_4Ni_3O_{10}$ samples from the same batches that exhibited superconducting $T_c$ under high pressure as reported in Refs. [8] and [13] respectively. For the $La_3Ni_2O_{7-\delta}$ sample, Laue pattern (Fig. 1B) confirms the a-b plane, and our SNOM study was conducted within the a-b plane. Figure 1C shows a typical near-field map $S(\mathbf{r})$ within a 20 um by 20 μm field of view collected on $La_3Ni_2O_{7-\delta}$ samples. A striking feature is the nanoscale separation of optical conductivity, characterized by enhanced stripy optical conductivity patterns. These conductivity stripes form domains, with their orientation 90 degree rotated between neighboring domains. We propose that the observed enhancement in the stripy conductivity is an intrinsic characteristic of the '327' phase of $La_3Ni_2O_{7-\delta}$, rather than arising from impurity structures such as $La_2NiO_4$[20], $La_4Ni_3O_{10}$[23], or even the '1313' structures typically present in $La_3Ni_2O_{7-\delta}$[21,57-59]. Firstly, the Laue diffraction pattern (Fig. 1B) reveals that these stripes run diagonally (x/y directions) relative to Ni-O-Ni crystal axes (a/b directions), whereas impurity phases typically align with the Ni-O-Ni directions, as observed in transmission electron microscope (TEM) studies[20-21,57]. Secondly, the same stripy features are evident in both AFM topography (Fig. 1D) and scanning electron microscope (SEM) images taken within the same field of view (SI Fig. 1C). As shown in Fig. 1E, step edges with a total height of ~15 nm on topography do not disturb conductivity stripes in the $S(\mathbf{r})$ map (Fig. 1G), indicating that these conductivity stripes are intrinsic bulk features of $La_3Ni_2O_{7-\delta}$ single crystals, rather than surface effects. Therefore, we concluded that the observed optical conductivity separation likely corresponds to phase separation in $La_3Ni_2O_{7-\delta}$ single crystal at ambient pressure, which will be discussed further in the following paragraph.

For comparison, we also performed SNOM measurements on the $La_4Ni_3O_{10}$ sample. As shown in Fig. 2 (and also Fig. S2), the $La_4Ni_3O_{10}$ sample exhibits uniform optical conductivity with a higher magnitude normalized with gold, but no evidence of phase separation, as revealed by the near-field signal map $S(\mathbf{r})$. The uniform optical conductivity observed in $La_4Ni_3O_{10}$ aligns with the nearly bulk superconductivity reported on the same batch under high pressure.

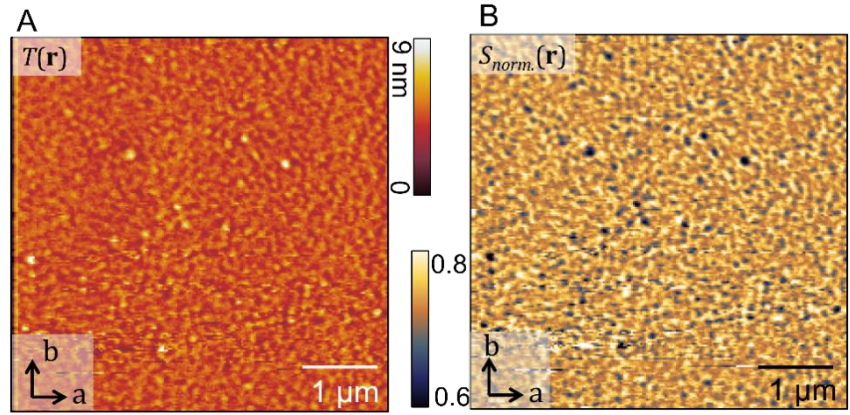

Figure. 2| **A**. AFM topography of La$_4$Ni$_3$O$_{10}$ crystals, $T(\mathbf{r})$. **B**. Simultaneously acquired near-field signal $S(\mathbf{r})$ of La$_4$Ni$_3$O$_{10}$ within the same FOV as $T(\mathbf{r})$ in Fig. A. See also Fig. S2. The $S(\mathbf{r})$ map shows uniform conductivity, with a magnitude of conductivity more than twice that of La$_3$Ni$_2$O$_{7-\delta}$. Note that the $S(\mathbf{r})$ was also normalized against the near-field signal measured on gold.

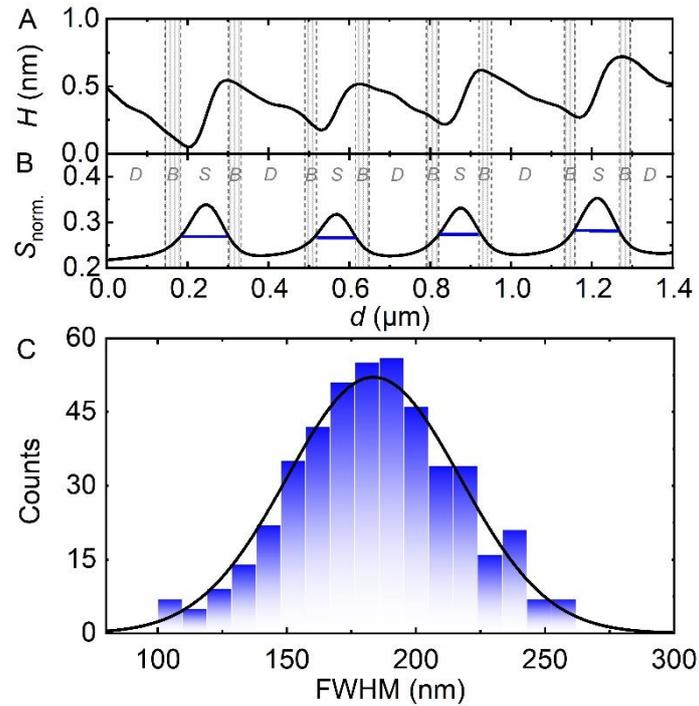

Figure. 3| **A,B**. Line cuts of topography $T(\mathbf{r})$ and near-field signal $S(\mathbf{r})$ of La$_3$Ni$_2$O$_7$ crossing multiple conductivity stripes along white solid line in Fig. 1B. Enhanced conductivity in $S(\mathbf{r})$ coincides with lower heights in the topography, $T(\mathbf{r})$. The blue bars mark full width at half maxima (FWHM) of each conductivity peak. Here the HWHM is used to represent the effective width of conductivity stripes. In this context, region D denotes the dark regions, region S denotes the bright stripes, and shaded region B represents the phase boundary area. Phase boundary regions are defined as the transitional areas where the optical conductivity gradually decreases from bright stripes to dark regions. **C**. Statistic histogram of FWHM values, counted from stripes measured from all La$_3$Ni$_2$O$_{7-\delta}$ samples. Black sold line represents the peak by Gaussian function with resultant peaked FWHM is ~183±33 nm.

To determine the length scale of optical conductivity stripes, Figures 3A and 3B show line profiles of topography T(r), and near-field signal $S(\mathbf{r})$ map taken perpendicular to the optical conductivity stripes, along white arrows on Figs. 1E and 1F. As depicted in Fig. 3A, the AFM topography reveals a zigzag-like variation in heights, indicative of the lattice mismatch between bright stripes and dark regions. It is evident that the peak and trough of the near-field signal are in consistent alignment with the steep and descending slopes of line profile of height. A similar phenomenon has been observed in the local conduction and topography of strained $BiFeO_3$ thin films, where phase separation occurs between the *R*-phase and *T'*-phase[60]. The existence of strain is frequently observed at the interface between different phases. This phenomenon can be attributed to the mismatch of lattice parameters. The line profiles were, thus, divided into three distinct regions. Region S was defined by bright stripes, region D by dark regions, and region B by the phase boundary area between former regions，which are clearly denoted in Figs. 3A and 3B. By counting the pixels in each region of Figs. 3A and 3B, the length ratio of bright stripes, dark regions and phase boundary area is approximately 36%, 22% and 44%, respectively. The full width at half maximum (FWHM) was used as a measure to the effective width of conductivity stripes. As indicated by blue bars in Fig. 3B, the FWHM of each stripe was successfully extracted using peak-fitting functions, yielding an effective stripe width of 183±33 nm. Notably, the MEP study[41] on $La_3Ni_2O_{7-\delta}$ revealed an oxygen stoichiometric stripe, which also measures of approximately 200 nm, suggesting a close relationship between the stripy conductivity enhancement and oxygen content in $La_3Ni_2O_{7-\delta}$. In conclusion, we have revealed nanoscale phase separation between dark reigons and bright stripes, with the bright stripes exhibiting higher optical conductivity.

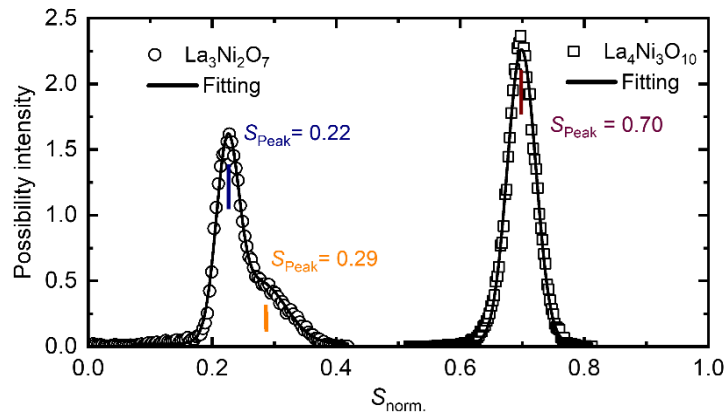

Figure. 4| Histograms of near-field signals for $La_3Ni_2O_{7-\delta}$ (hollow circle) and $La_4Ni_3O_{10}$ (hollow square) are shown within the same FOV size. Horizontal axis represents the normalized near-field signals to gold, $S_{\text{norm.}} = \frac{S_{\text{sample}}}{S_{\text{gold}}}$. Vertical axis shows possibility intensity of histograms calculated by $\int \rho_S \, dS = 1$. For the histogram of $La_3Ni_2O_{7-\delta}$, there is one main peak and a relatively broad side peak, which originates from the phase separation between dark regions and bright stripes. Solid lines represent fitting results (SI Note). The main peak (blue bar) from the dark regions has a peak position of $S_{peak}$~0.22 (22% of conductivity on gold), while the side peak (orange bar) with a peak position $S_{peak}$~0.29 (29% of conductivity on gold) corresponds with conductivity stripes of bright stripes. Intriguingly, the bright stripes constitute approximately 38 % of the total FOV, closely matching the superconducting fraction under high pressure. In comparison, $La_4Ni_3O_{10}$

exhibits uniform conductivity (brown bar) with a higher magnitude, $S_{peak}$~0.7 (70% of conductivity on gold).

To quantify the optical conductivity of La$_3$Ni$_2$O$_{7-\delta}$ and La$_4$Ni$_3$O$_{10}$, we normalized the near-field $S(\mathbf{r})$ maps by the magnitude of gold through $S_{\text{norm.}} = \frac{S_{\text{sample}}}{S_{\text{gold}}}$. Histograms of the normalized near-field signal $S(\mathbf{r})$ representing optical conductivity are shown in Fig. 4. For La$_3$Ni$_2$O$_{7-\delta}$, two conductivity peaks of 22% and 29% of gold's conductivity are associated with dark regions and bright stripes, respectively. We found that the optical conductivity measured by SNOM at the nanoscale is overall consistent with the electronic conductivity measured in transport measurements[39-40]. The bright stripes in Fig. 1B, accounts for 38% of the field of view (Referring to SI Notes for method), while the remaining 62% consists of dark and boundary regions. In contrast, La$_4$Ni$_3$O$_{10}$ exhibits a single optical conductivity peak at approximately 70% of that of gold.

**Discussion**

We propose that the observed optical conductivity stripes likely originates from structural phase separation. Neutron powder diffraction studies conducted on polycrystalline La$_3$Ni$_2$O$_{7-\delta}$ powder samples, as detailed in Ref. [61], confirmed the presence of both *Amam* and *Fmmm* crystal at ambient pressure, in agreement with early findings [36]. The refined compositions for the two phases are 56.1% La$_3$Ni$_2$O$_{6.79}$ (δ=0.21) for the *Amam* structure and 41.0% La$_3$Ni$_2$O$_{6.90}$ (δ = 0.1) for the *Fmmm* structure[61]. This ratio observed by neutron powder diffraction studies is comparable to that of dark regions and bright stripes. In the high pressure *Fmmm* structure, one γ-band contributed by $3d_{z^2}$ orbitals, emerges above the Fermi level, contributing a higher low-energy density of states compared to the *Amam* structure[8,32]. Consequently, the *Fmmm* structure, with fewer oxygen deficiencies[61], demonstrates higher conductivity compared to the *Amam* structure.

The structural phase separation revealed in this study is reminiscent of the phase separation in the discovered in the iron-based superconductor K$_x$Fe$_{2-y}$Se$_2$, where the '122' phase is widely believed to be the superconducting one, while other phases, such as '245' with iron vacancies, are magnetic and insulating[62-64]. Similarly, in La$_3$Ni$_2$O$_{7-\delta}$, it is crucial to explore how nanoscale structural phase separation contributes to the understanding of magnetic excitations, as observed in RIXS, NMR and the Neutron experiments[36,61,65-67]. This comparison may shed light on the role of phase separation in determining both superconducting and magnetic properties in these materials.

The low superconducting volume fraction[20-21] has sparked significant debate over whether superconductivity in La$_3$Ni$_2$O$_{7-\delta}$ is filamentary or indicative of a bulk nature. Zhou *et al.*[20] suggested that the low superconducting volume fraction in single crystal may result from a mixture of La$_4$Ni$_3$O$_{10}$ and La$_3$Ni$_2$O$_{7-\delta}$ phases. Wang *et al.*[22] demonstrated that partially doping Pr into La sites, La$_2$PrNi$_2$O$_{7-\delta}$ polycrystalline, reduces the proportion of the tri-layer La$_4$Ni$_3$O$_{10}$ phase, leading to an increase of the superconducting volume fraction above 50%. In this study, the revealed nanoscale structural phase separation at ambient pressure offers a new perspective on the possible structural origin of high temperature superconductivity in La$_3$Ni$_2$O$_{7-\delta}$ single crystal. Assuming that one structure within the La$_3$Ni$_2$O$_{7-\delta}$ crystal becomes superconducting under high pressure, the obtained area ratio of bright stripes and residual

(including dark and boundary regions) close to 4:6 aligns with the reported superconducting fraction ~57% in $La_2PrNi_2O_{7-\delta}$ under high pressure[22]. This correlation suggests that the high-temperature superconductivity may emerge firstly from one structure at certain high pressure, and that interruptions by other structures may be responsible for the reduced superconducting volume fraction. Huo *et al.*[17] reported the signature of superconductivity with $T_c$ ~80 K at ambient pressure in bulk sample $La_3Ni_2O_7$, in which the superconducting volume fraction is estimated to be within 0.2%. The superconductivity at ambient pressure with extremely low superconducting volume fraction may also be associated with the nanoscale structural phase separation observed in our experiments. In contrast, $La_4Ni_3O_{10}$ exhibits uniform optical conductivity, more than twice that of $La_3Ni_2O_{7-\delta}$, with no evidence of phase separation. This is consistent with the nearly bulk superconductivity reported in $La_4Ni_3O_{10}$ under high pressure[13].

**Conclusion**

We conducted SNOM studies on the nickelate compounds $La_{n+1}Ni_nO_{3n+1}$ (n=2,3) to investigate low-energy physics at the nanoscale. By imaging local optical conductivities, we observed nanoscale structural phase separation in $La_3Ni_2O_{7-\delta}$, with local conductivity values of 22% and 29% of that of gold, respectively. Our findings suggest that the bulk superconductivity observed in $La_3Ni_2O_{7-\delta}$ is associated with one structure at ambient pressure, while interruptions by other regions likely contribute to the low superconducting volume fraction under high pressure. In contrast, $La_4Ni_3O_{10}$ exhibits uniform optical conductivity with a magnitude ~70% of that on gold. Upon the application of pressure, a structure transition from orthorhombic to tetragonal *I4/mmm* phase likely occurs in both $La_3Ni_2O_{7-\delta}$ and $La_4Ni_3O_{10}$, with superconductivity emerging in the tetragonal phase at low temperatures. In conclusion, our SNOM study on $La_{n+1}Ni_nO_{3n+1}$ (n=2,3) provides new insights into the structural origin of high-temperature superconductivity in nickelate compounds, highlighting SNOM as a crucial technique for probing nanoscale phenomena in correlated materials.

**Methods**
**Sample synthesis:** Both $La_3Ni_2O_{7-\delta}$ and $La_4Ni_3O_{10}$ samples were fabricated by the high oxygen pressure floating zone technique and the details are described in Refs. [8,13,68]. $La_3Ni_2O_{7-\delta}$ sample was checked by X-ray diffraction (Fig. S1A) and Laue diffraction (Fig. 1B). All samples were cleaved to get a flat, clean surface before SNOM, SEM measurements.
**s-SNOM experiments**: SNOM measurements were conducted on a commercial NeaSpec system at Anhui University, employing a QCL laser centered at ~8 um at room temperature. The laser power before shining onto to the AFM tip was around 25 mW and it was reduced to below 5 mW at the tip. The near-field signal is firstly aligned using a deposited gold thin film as reference, which was placed alongside the $La_3Ni_2O_{7-\delta}$ sample. Then, the tip was moved on to the samples to collect the near-field signal. We utilized a heterodyne detection scheme, which provides stable detection of the scattered near-field signal and ensures signal contrast. All near-field maps presented in this study are obtained in the third harmonic.

**Acknowledgement**

We thank Mengkun Liu, Tao Wu, Zhengyu Wang, Shoucong Ning, Yangmu Li for helpful discussion and Lukas Wehmeier, Xinzhong Chen for assistance during experiments. D.L.F acknowledges the support by the New Cornerstone Science Foundation (No. NCI202211), and the Innovation Program for Quantum Science and Technology (No. 2021ZD0302803). Z.D acknowledges the support by Hefei National Laboratory. Y.Z., E,Z, and J.Z. were supported by the National Key R&D Program of China (Grant No. 2022YFA1403202), the Key Program of National Natural Science Foundation of China (Grant No. 12234006), and Innovation Program for Quantum Science and Technology (Grant No. 2024ZD0300103). Work at SYSU was supported by the Natural Science Foundation of China (Grant No. 12174454), the National Key Research and Development Program of China (Grant No. 2023YFA1406500), the Guangdong Basic and Applied Basic Research Funds (Grant No. 2024B1515020040), Guangzhou Basic and Applied Basic Research Funds (Grant No. 2024A04J6417), and Guangdong Provincial Key Laboratory of Magnetoelectric Physics and Devices (Grant No. 2022B1212010008).